\documentclass[11pt,fleqn]{article}
\usepackage[utf8]{inputenc}
\usepackage{fullpage}

\usepackage{amsthm}
\usepackage[dvipsnames]{xcolor}
\usepackage{hyperref}
\usepackage{todonotes}
\usepackage{qcircuit}
\usepackage{cite}
\usepackage{tikz}
\usepackage{tikzsymbols}
\usepackage{titling}
\usepackage{textcomp}
\usepackage{mdframed}

\usepackage{libertine}
\usepackage{libertinust1math}
\usepackage[T1]{fontenc}

\usepackage{tikz}
\usepackage{pifont}
\usepackage{bbding}

\usepackage{tikz}
\usetikzlibrary{shapes,backgrounds}

\usepackage{complexity}

\usepackage{setspace}
\usepackage{nirkhe}

\allowdisplaybreaks 

\usepackage{tikz}
\usetikzlibrary{shapes,backgrounds}

\usepackage{cleveref}

\newtheorem{theorem}{Theorem}

    \crefname{definition}{definition}{definitions}
\newtheorem{lemma}[theorem]{Lemma}
    \crefname{lemma}{lemma}{lemmas}
\newtheorem{conjecture}[theorem]{Conjecture}
    \crefname{conjecture}{conjecture}{conjectures}

    \crefname{corollary}{corollary}{corollaries}

    \crefname{fact}{fact}{facts}
\newtheorem{remark}[theorem]{Remark}
    \crefname{remark}{remark}{remarks}

\newenvironment{xalign}[1][]{
    \subequations
    \label{#1}
    \align
}
{
    \endalign
    \endsubequations
}

\makeatletter
\renewcommand{\maketitle}{\bgroup\setlength{\parindent}{0pt}
\begin{flushleft}
  \LARGE {\@title} \linebreak
  
  \normalsize \@author \linebreak
\end{flushleft}\egroup
}
\makeatother

\usepackage{authblk}

\title{The status of the quantum PCP conjecture (games version)}

\author[1]{Anand Natarajan}
\author[2]{Chinmay Nirkhe}

\affil[1]{\footnotesize Massachusetts Institute of Technology,  \it{Cambridge, Mass.}}
\affil[2]{\footnotesize IBM Quantum, Thomas J. Watson Research Center, \it{Yorktown Heights, N.Y.}}

\date{}

\begin{document}

\maketitle{}

\begin{abstract}
    In classical complexity theory, the two definitions of probabilistically checkable proofs -- the constraint satisfaction and the nonlocal games version -- are computationally equal in power. In the quantum setting, the situation is far less clear. The result $\MIP^* = \RE$ of Ji \emph{et. al.}~\cite{2001.04383} and refinements by Natarajan and Zhang~\cite{quantum-free-games} show that multiprover interactive proof systems with polylogarithmically long messages can solve any decision problem in $\RE$, including undecidable problems like the halting problem. These results show that any connection between the ``constraint satisfaction'' or ``Hamiltonian'' quantum PCP conjecture and nonlocal games must involve restricting the players in the game to be \emph{computationally efficient}. This note contains two main results: (1) we give a ``quantum games PCP for $\AM$'' in the form of a new construction of a succinct $\MIP^*$ protocol with efficient provers for the canonical $\AM$-complete problem, and (2) we explain an error in the energy amplification procedure of Natarajan and Vidick~\cite{errata-games-qpcp} which invalidates their claim to have constructed a quantum games PCP for a $\QMA$-complete problem. In surveying the obstacles remaining towards a quantum games PCP for $\QMA$, we highlight the importance and challenge of understanding gap amplification for Hamiltonians even when locality is replaced by much weaker constraints, such as bounds on the ``Pauli spectrum'' of the Hamiltonian. We hope these questions will motivate progress towards new ``baby versions'' of Hamiltonian quantum PCP conjecture.
\end{abstract}

\section{Introduction}

\subsection{Interactive proofs with entanglement}

How powerful is an interactive proof system with provers sharing quantum entanglement? In 2020, Ji, Natarajan, Vidick, Wright, and Yuen proved that such an interactive proof system can be used to decide all recursively enumerable languages \cite{2001.04383}. Equivalently, they proved the equality $\MIP^* = \RE$ between the respective complexity classes.

More specifically, they showed that $\RE$ is captured by \emph{nonlocal games}, which can be thought of as an interaction between a verifier and two arbitrarily powerful devices, often denoted as Alice and Bob. Alice and Bob are spatially separated but allowed to share entanglement. The verifier (using private randomness) samples questions for Alice and Bob who then answer using their shared entanglement. The verifier then evaluates a binary relation on the question and answer pairs, and accepts or rejects accordingly.

A direct consequence of the $\MIP^* = \RE$ result \cite{2001.04383} is the construction of a protocol with the following properties. A verifier --- whose input is the description of a Turing machine, $\langle T \rangle$, and intention is to decide whether $T$ halts or not --- sends messages $q_1, q_2$ to entangled and non-communicating Alice and Bob, respectively. The verifier receives answers $a_1, a_2$, respectively, and then performs a randomized computation $V_T(q_1, q_2, a_1, a_2)$. The property of the protocol is that there exists a strategy for Alice and Bob which causes the verifier to accept with probability 1 if $T$ halts while \emph{all} strategies for Alice and Bob cause the verifier to reject with high probability if $T$ does not halt. Crucially, the algorithm $V_T$ has a runtime which is $\poly(n)$ where $n = \abs{\langle T \rangle}$ is the length of the description of the Turing machine; this implies that the questions and answers also have length at most $\poly(n)$. While the complexity of the verifier must be at least linear in $n$ (by standard arguments), can the communication be shortened? More specifically, can the questions and answer lengths be made shorter?

The answer is -- surprisingly -- yes! Natarajan and Zhang \cite{quantum-free-games} improved the $\MIP^* = \RE$ result \cite{2001.04383} to prove that estimating the entangled value of a nonlocal game with either questions or answers of length $\poly \log(n)$ captures the $\RE$ complete problem of deciding if a Turing machine $T$ halts. This result can be seen as partial progress towards resolving the \emph{quantum games PCP conjecture} where PCP stands for Probabalistically Checkable Proofs. We will state this conjecture more precisely below in \Cref{sec:towards-qpcp}, but first, to put the result in context, let us take a detour to the classical PCP theorem, and the complexity of classical nonlocal games.

\subsection{Probabilistically Checkable Proofs and Games}

The power of nonlocal games---or multiprover interactive proof systems---with classical, unentangled players, was exactly characterized in a sequence of results~\cite{bfl90,feige_et_al,Arora:1998:PCP:273865.273901,Haastad-pcp}\footnote{The sequence of results stems from a result by Babai, Fortnow, and Lund \cite{bfl90} which showed a multiround interaction between multiple provers and a verifier for $\NEXP$ with $O(\poly(n))$ sized total question length and $O(\poly(n))$ sized total answer length. Later results \cite{feige_et_al,Arora:1998:PCP:273865.273901,Haastad-pcp} reduced the round complexity to 1, answer length to $O(1)$, and demonstrated a compression which gave the stated equivalence to $\NP$.} leading up to the PCP theorem. This theorem has several equivalent formulations, but in terms of multiprover interactive proofs (or more colloquially, nonlocal games), it states that the class $\NP$ is exactly equal to the class of problems that can be decided by one-round, two-prover $\MIP$ proof systems with $O(\log n)$-length questions and $O(1)$-length answers.

Note the following two key points of the classical games PCP theorem. Firstly, the parameters for the question and answer length immediately yield the more familiar ``probabilistically checkable proofs" version of the PCP theorem. This is because any deterministic strategy to a 2-prover nonlocal game with length-$q$ questions and length-$a$ answers can be written down in a table of size $O(2^q \cdot a)$, and the interaction between the provers and verifier can be simulated by querying $2a$ bits in the table. This means that for any language in $\NP$, there is a \emph{probabilistically checkable proof system} for it: the verifier receives a string of length $O(2^q \cdot a ) = \poly(n)$ (which in the honest case is the table corresponding to the optimal prover strategy), and makes $2a = O(1)$ queries to it. If the answer is YES, then the verifier will accept the honest proof string with high probability, whereas if the answer is NO, it will reject all proof strings with high probability. In fact, the implication goes the other way as well: any probabilistically checkable proof with polynomial-sized proofs implies a two-player interactive proof system with $O(\log n)$-length questions and $O(1)$-length answers. This is due to a standard transformation called the \emph{clause-variable game}; a detailed treatment can be found in Thomas Vidick's lecture notes on the quantum PCP conjecture~\cite{vidick-qpcp-lecturenotes}. Thus, the two formulations of the classical PCP theorem are equivalent. 

Secondly, the protocol arising from the games version of the PCP theorem is \emph{prover efficient}\footnote{There is also a natural concern of the efficiency of a verifier. In some context a proof that is both verifier and prover efficient is called doubly efficient. However, in most such cases, such as the Goldwasser, Kalai, and Rothblum interactive proofs \cite{gkr}, the notion of efficiency is of a fine grained nature concerning the efficiency of the verifer and prover in terms of the size or depth of the formula. Specifically, for \cite{gkr}, the verifier should run in time $O(n \poly(d))$ and space $O(\log n)$ and the prover in time $\poly(n)$ where $d$ is the depth of the circuit. However, in this note we are focused on the coarser perspective on complexity and therefore we are content with the verifier being efficient if they are a $\BPP$ or $\BQP$ device. Therefore, to emphasize the importance on making the prover efficient, we call these problems only prover efficient and not doubly efficient.}. In the classical nonlocal games setting, we find that in reduction from any language $\Ll \in \NP$ to a family of nonlocal games, the \emph{honest} classical provers in the game can generate (in polynomial time) the winning answers to the questions from the witness $w$ to the original $\NP$ language. In other words, the honest provers only need to be $\P^w$ powerful. Meanwhile, the protocol is sound against arbitrarily powerful classical powers.

\subsection{The quantum PCP conjectures}

In the quantum case, we have no PCP theorem, but rather several formulations of quantum PCP conjectures, which are not known to be equivalent. A standard version of the conjecture is the ``Hamiltonian qPCP conjecture", which states that a gapped version of the local Hamiltonian problem is $\QMA$-complete under quantum polynomial-time reductions. 
\begin{conjecture}[Quantum PCP \cite{QuantumNPsurvey,10.1145/2491533.2491549}]
It is $\QMA$-complete under quantum polynomial-time reductions to decide\footnote{The constants of $1/10$ and $1/5$ could be replaced with any other choice of constants.} whether a local Hamiltonian on $n$ qubits and $m = \Theta(n)$ terms has ground energy (minimum eigenvalue) $\leq m/10$ (YES instance) or $\geq m/5$ (NO instance) even when promised that one of the cases holds.
\end{conjecture}
This conjecture can equivalently be stated in terms of the existence of probabilistically checkable \emph{quantum} proofs for any language in $\QMA$. This conjecture is discussed extensively in the survey by Aharonov, Arad, and Vidick~\cite{10.1145/2491533.2491549}. Another version of the conjecture concerns the ``proof-checking'' property of a quantum PCP. It is known that the proof-checking version of the conjecture is equivalent ot the Hamiltonian version under quantum polynomial-time reductions since the previously mentioned survey~\cite{10.1145/2491533.2491549}. A recent note by Burhman, Helsen, and Weggemans \cite{buhrman2024quantum} has resolved certain questions about the \emph{adaptivity} of the proof-checking and connections to the complexity class $\QCMA$ -- however, we do not discuss the proof-checking version any further in this note.

In addition to trying to generalize the proof-checking version of the classical PCP, attempts have been made to formulate a \emph{quantum games PCP}, generalizing the statement of the classical PCP theorem in terms of games or $\MIP$ proof systems for $\NP$. This direction was first proposed by Fitzsimons and Vidick~\cite{10.1145/2688073.2688094}, but in light of the subsequent progress in our understanding of $\MIP^*$, it is worth revisiting it. Motivated by our presentation of the classical games PCP, we propose that a reasonable statement of a quantum games PCP conjecture should satisfy the following requirements:
\begin{itemize}
    \item It should put $\QMA$ into a class of $\MIP^*$ proof systems with short questions and answers. 
    \item The $\MIP^*$ proof system should have question length $q = O(\log n)$ and answer length $a = O(1)$. 
    \item Honest provers should be efficient, given (copies of) the $\QMA$ witness.
\end{itemize}
The purpose of these requirements is not to blindly mimic the classical case, but to preserve the hope that the resulting $\MIP^*$ proof system will say something about the Hamiltonian qPCP conjecture. In particular, one may hope that these constraints will result in a proof system where the honest provers' strategy involves constructing some kind of polynomial-sized quantum probabilistically checkable proof, by performing an efficient transformation on the $\QMA$ witness.

\subsection{Towards a quantum games PCP}
\label{sec:towards-qpcp}
We previously remarked that the Natarajan and Zhang result \cite{quantum-free-games} is only partial progress towards a quantum games PCP. This is because it fails to satisfy the second and third requirements given above (and arguably only satisfies the first requirement by a technicality, since the result applies to all of $\RE$ and not just $\QMA$).

The fact that the Natarajan and Zhang result does not achieve the gold standard of $O(\log n)$ sized questions and $O(1)$ sized answers, and instead requires $\poly \log(n)$ sized questions and answers, is the more minor of two reasons. We believe the roadblock to be more minor as the original games PCP for $\NP$ by Feige \emph{et. al.} \cite{feige_et_al} involved $O(\poly\log n)$ length messages and subsequent improvements yielded $O(\log n)$ length messages \cite{Arora:1998:PCP:273865.273901,Arora:1998:PVH:278298.278306}. It seems plausible that similar improvements can be achieved in the $\MIP^*$ setting by refining known techniques (although it is worth noting that, as remarked in \cite{quantum-free-games}, achieving this improvement will require improving or replacing the analysis of the quantum soundess of the low-degree test in \cite{Ji2020QuantumSO}).

The second -- and more substantial roadblock -- is the complexity of the \emph{honest} provers. The quantum analog of the efficient-provers property is to consider scenarios where the honest prover is not all-powerful but rather is only $\BQP$-powerful but imbued with the solution to the particular problem. For example, in the case of a $\NP$ problem $x$, we can morally think of the honest provers, by analogy, as $\BQP^w$ powerful where $w$ is the witness to the problem. More technically, the provers will need to share entanglement and potentially this entanglement may rely on the witness $w$. Recall that a nonlocal game consists of three phases: (a) a setup phase where Alice and Bob generate their specific entangled state $\rho_{AB}$, (b) an interaction phase where the game is played with the verifier, (c) and a grading phase done exclusively by the verifier. The technical definition of efficient provers will be one where (a) the setup phase can be performed by a quantum polynomial-time (QPT) device with access to $w$ and (b) the interaction phases can be performed by QPT devices with access to their respective share of the state $\rho_{AB}$.
To extend this definition to $\QMA$, a class where the witness is quantum, we define an efficient nonlocal game capturing a language in $\QMA$ like the $\NP$ definition except the classical proof $w$ is replaced by $\ket{\psi}^{\otimes \poly(n)}$ in the setup phase, where $\ket{\psi}$ is the quantum witness for the problem. The rationale for multiple copies of the witness is that the efficient provers cannot (in general) clone $\ket{\psi}$, and many standard reductions between $\QMA$ protocols (e.g. amplification procedures)  require multiple copies of the witness.

It is important to emphasize that prover efficiency is a property of a reduction from one language to another and not a property of a complexity class. This is because complexity class equalities and reductions between languages within a complexity class may not be prover efficient. The most pertinent example to keep in mind is that, even though $\QMA \subseteq \NEXP$, and therefore, the $\QMA$-complete local Hamiltonian problem can be reduced to a $\NEXP$-complete problem such as Succinct-3-Coloring, the witness for the coloring problem is likely not efficiently computable from the quantum witness to the local Hamiltonian problem.
Therefore, when discussing prover efficiency we will be careful to specify a language (and not a complexity class) and a specific model of the proof.

A priori, one might suspect that the construction of prover-efficient nonlocal games with short questions would be easiest for a language in the class $\QMA$ where the witness is a quantum state since the state of the honest provers could simply be the witness for the $\QMA$ problem. However, as we remark in this note, there are significant roadblocks to constructing such games and a construction is not known. We believe that this is the most practical question remaining in the pantheon of nonlocal game theory and, therefore, the appropriate question to be called the quantum PCP games conjecture.

\begin{conjecture}[Quantum PCP (Games Version)]
There exists a prover-efficient $\MIP^*$ protocol with $O(\log n)$-length questions and $O(1)$-length answers for $\QMA$. \\

\noindent More formally, for every language $\Ll \in \QMA$, there exists a polynomial time verifier $V$ and quantum polynomial time provers $P_1, \dots, P_k$ such that:
\begin{enumerate}
\item If $x \in \Ll$, then there exists a $\poly(n)$-qubit state $\ket{\psi}$ such that $V$ on input $x$, interacting with $P_1, \dots, P_k$ on input $(x, \ket{\psi}^{\otimes \poly (n)})$, accepts with probability $2/3$. In particular, the provers $P_1, \dots, P_k$, in the setup phase of the protocol, generate their shared entangled state in polynomial time from the input $(x, \ket{\psi}^{\otimes \poly(n)})$.
    \item If $x \not\in \Ll$, then for any (not necessarily efficient) provers $P^*_1, \dots, P^*_k$, the verifier $V$ given input $x$ and interacting with $P^*_1, \dots, P^*_k$ accepts with probability at most $1/3$.
\end{enumerate}

We remark that here, $\ket{\psi}$ is allowed to be an arbitrary state depending on $x$. However, one may think of it as a $\QMA$ witness for $x$: indeed, any $\MIP^*$ protocol of this form implies a $\QMA$ protocol for $\Ll$, where honest witness is $\ket{\psi}^{\otimes n}$, and the $\QMA$ verifier simulates the interaction between $V$ and $P_1, \dots, P_k$ (since the provers are efficient). 

\end{conjecture}

The $\MIP^* = \RE$ result of~\cite{2001.04383} does not directly say anything about the efficiency of the provers, but by inspecting the strategy for the honest provers given in the completeness case of the protocol, we can obtain explicit upper bounds on the prover runtime when the language $\Ll$ to be decided is in a time-bounded complexity class. Unfortunately, these bounds are not good enough for $\QMA$, because they only depend on the \emph{classical} nondeterministic time complexity of $\Ll$. Specifically, for any problem in $\NTIME[t(n)]$, the honest prover strategy requires quantum time $\poly(t(n))$, given a classical nondeterministic witness of size $t(n)$. This strategy requires the prover to compute a PCP-style encoding $\pi$ of a classical tableau of the execution of a Turing machine solving the problem in $t(n)$ steps, and prepare on the order of $\poly\log t(n)$ EPR pairs; in response to a verifier question, the prover measures the EPR pairs to obtain indices into the encoded tableau, and reports the value of the tableau at those indices. For the specific case of $\BQP$ or $\QMA$, the best known $\NTIME[t(n)]$ bounds on these classes are $t(n) = \exp(n)$. Moreover, for $\QMA$, even given copies of the witness state, we do not know how an quantum prover could compute entries of $\pi$ in time less than $t(n)$.

\subsection{Contributions of this note}

In this note, we prove two results and comment on the current state of affairs. The first is that prover-efficient nonlocal games exist for $\AM$ protocols with $\poly \log (n)$ sized questions and answers. Second, we describe the error in the incorrect result in Natarajan and Vidick \cite{errata-games-qpcp} claiming a reduction from the local Hamiltonian problems to quantum games PCPs.  %
Lastly, we remark on the outstanding roadblocks for the games version of the QPCP conjecture -- i.e. constructing prover-efficient nonlocal games for $\QMA$. 

\section{Preliminaries}

\subsection{Nomenclature and notation}

The majority of this note regards the computational complexity of computing the entangled value of a nonlocal game or the $\QMA$-complete problem of computing the minimum eigenvalue of a local Hamiltonian. To simplify the reductions between these two problems, we will instead consider computing the \emph{maximum} eigenvalue of a local Hamiltonian.

The problems can be expressed as optimization problems, but for connections to the pantheon of computational complexity classes, it is useful to consider the \emph{promise-gapped} version of the decision problem. The problem of deciding, for a game $G$ whether the entangled value, $\omega^*(G)$, is $\geq$ than $c$ (YES instances) or $\leq s$ (NO instances) is the promise-gapped version of the problem with $\pgap = c - s$. Likewise, the promise-gapped version of the local Hamiltonian problem is to decide for a local Hamiltonian $\H$, whether (YES instances) $\lambda_{\max}(\H) \geq c$ or (NO instances) $\lambda_{\max}(\H) \leq s$. 

The principle goal of this note is to perform reductions between games and Hamiltonians that amplify the promise gap. Notationally, we will discuss ``amplifying the promise gap'' of the game (or the Hamiltonian) when we precisely mean generate a reduction from a set of games (or Hamiltonian instances) to a new set of games (or Hamiltonian instances) with a promise gap between YES and NO instances greater than the promise gap of the initial set. Lastly, unless specified otherwise, when discussing games we are always referring to the decision problem of estimating the entangled value of the game.

\subsection{The computational power of prover \emph{non-efficient} nonlocal games}

As previously mentioned in the Introduction, stemming from the introspection (i.e. question and answer reduction) tools developed in \cite{natarajan_wright_2019,2001.04383}, Natarajan and Zhang \cite{quantum-free-games} proved that any Turing machine halting question can be transformed into a nonlocal game. This is the \emph{state-of-the-art} in terms of succinct nonlocal games. Note, the following theorem states nothing about the efficiency of the honest provers.

\begin{theorem}[Theorem 58 in \cite{quantum-free-games}]
$\MIP^*[q = O(1), a = O(\poly\log n)] = \RE$. 
\end{theorem}

The appropriate interpretation is the every Turing machine halting problem, defined by input $\ev{T}$ with $n \defeq \abs{\ev{T}}$ can be reducing to deciding the quantum value $\omega^\star$ of a nonlocal game described by a table with constant sized questions and $O(\poly\log n)$ sized answers -- i.e. the tableau of the verification function\footnote{This table has size $O(4^{q + a}) = \textrm{quasipoly}(n)$.} $V(q_1, q_2, a_1, a_2)$. Moreover, the table of the game can be computed from the description of the Turing machine in classical \emph{quasi-polynomial time}. Equivalently, $\MIP^\star \supseteq \RE$ under quasi-polynomial time reductions since the Turing machine halting problem is $\RE$-complete. The containment $\MIP^\star \subseteq \RE$ under quasi-polynomial time reductions is straightforward. 

The nuance of the $O(\poly \log n)$ sized questions versus the $O(\log n)$ sized questions in the classical games PCP stems from $O(\poly \log n)$ sized questions required for the \emph{quantum low individual degree test} \cite{Ji2020QuantumSO} which is a seminal step in the introspection arguments used in \cite{2001.04383} and \cite{quantum-free-games}. We believe (but it deserves a closer inspection through the entire lengthy proof) that a more efficient alternative to the quantum low individual degree test would close the gap on the polylogarithmic vs logarithmic difference between the quantum and classical variants.

\section{An efficient nonlocal game for all $\AM$ languages}

An $\AM$ protocol for deciding language membership $x \in \Ll$ is an interaction between a prover (Merlin) and a public-key verifier (Arthur) where in the first message Arthur sends a uniformly random $r \in \bits^{\poly(n)} = R$ and then Merlin responds with $w(r)$ according to a witness $\fn{w}{R}{\bits^{\poly(n)}}$. Arthur follows by running a deterministic computation $V(x, r, w(r))$.

It is well known that $\AM$ protocols can be constructed with completeness $1$ and soundness $\leq 1/3$. We can call an honest prover efficient (or for brevity, efficient) if the honest prover can be simulated by a $\P^{w}$ machine -- i.e. a polynomial-time machine which has access to $w$ but no other ``super-natural'' computational abilities. In the entangled game setting, we are interested in understand efficient nonlocal games where the honest provers can be simulated by $\BQP^{w}$ machines. The amplification from $\P$ to $\BQP$ is inherently necessary as the provers must share entanglement in order to have more computational power than $\NP$ --- and as we don't know if $\AM = \NP$, we must consider entangled strategies.

To understand the efficient nonlocal game for $\AM$ at a high level, consider the following honest provers, Alice and Bob. They receive no input from the verifier, and instead sample their own questions by measuring $\abs{r}$ EPR pairs in the same basis to generate the same question string $r$. Alice's answer to the verifier is simply $r$, and Bob's answer to the verifier is $(r, w(r))$. The verifier then checks that the answers have consistent strings $r$ and that $V(x,r,w(r))$ accepts. 

Notice, that if the randomness $r$ is sampled correctly, then by the soundness of the original $\AM$ problem, the provers are incapable of any deceit as they must answer a witness for $r$. Of course, the principal issue is that we cannot trust nefarious provers to sample their own questions. This is where we can use the introspection tools of \cite{2001.04383}. With introspection, a verifier can use $\poly\log(n)$ sized questions and force the provers to sample their own questions. This gives us a game with short questions and long answers for $\AM$.

The next step is to have the provers grade their own witness --- or equivalently, answer-reduction. Since Bob's proof has both $r$ and $w$ and $x$ is public, Bob can generate a \emph{probabilistically checkable proof of proximity} (PCPP) of the fact that $V(x,r,w)$ accepts. Applying this intuition into the answer reduction argument of \cite{2001.04383} gets us short questions and answers.

\begin{theorem}
    For every language $\Ll \in \AM$, there is a reduction from $\Ll$ to a language of nonlocal games with questions and answers of size $O(\polylog(n))$ such that (a) there is a reduction from problems in $\Ll$ to the entangled value of the game and (b) for any YES instance of the language $\Ll$ with witness $w= w(r)$, the honest provers only need be $\BQP$-powerful with knowledge of $w$.  
    \label{thm:AM}
\end{theorem}

Furthermore, one might ask whether prover-efficient succinct nonlocal games are achievable for languages in classes significantly larger than $\AM$ or $\QMA$, such as $\NEXP$, $\NEEXP$, or even $\RE$. While one can make such claims, the issue -- as we suggested earlier -- is in the subtelties of the definition of prover efficiency. 
Consider, for instance, a reduction from the $\NEXP$-complete language of Succinct-3-Coloring to a nonlocal game. A consequence of Impagliazzo, Kabanets, and Wigderson \cite{IMPAGLIAZZO2002672} shows that unless $\NEXP \subseteq \Sigma_2$, the coloring function for Succinct-3-Coloring problems cannot be expressed succinctly. Therefore, assuming this monumental complexity class collapse is false, we can only prove prover efficiency in a model of \emph{oracle access} to the witness instead of the prover actually holding the witness itself\footnote{An astute reader might find this argument peculiar since we considered provers with access to the $\AM$ witness function $w(r)$ in the previous example. Indeed, if the function $w$ were efficiently describable, the problem is actually in the class $\MA$ as the prover could send the description of the function $w$, upon which the verifier could sample randomness and verify that $w \cong 1$. The principle difference in these two examples is that the derandomization of $\AM = \MA = \NP$ is well believed to be true, while the collapse of the polynomial hierarchy in $\NEXP \subseteq \AM$ is widely believed to be false.}. 

Secondly, its worth noting that we know of prover-efficient nonlocal games for the class $\PCP[q = \poly(n), a = O(1)]$ as the original proof of $\MIP^\star \subset \MIP$ by Ito and Vidick \cite{6375302} is prover efficient. However, the transformation from a witness of a $\NEXP$-complete language such as Succinct-3-Coloring to one for the $\PCP$ equivalent language may not be prover efficient. For this to be the case, we need that the witness $w'$ for the PCP version of the instance can be simulated with polynomial queries to the witness $w$ of the $\NEXP$ version of the instance. To the best of our knowledge, applications of the known PCP transformations such as that of Dinur \cite{dinur-pcp} require exponentially many queries to $w$. While we remark, but do not explicitly prove here, that a similar proof to the $\AM$ argument above yields a quantum games PCP for the canonical language for $\PCP[q = \poly(n), a = O(1)]$ that is prover efficient in the sense that the prover has oracle access to the witness for the PCP, we want to emphasize that this doesn't necessarily yield a prover-efficient argument for all $\NEXP$ languages. A similar remark extends for languages in complexity classes past $\NEXP$.

\section{The difficulty in constructing nonlocal games for languages in $\QMA$}

Given that $\MIP^*$ allows for quantum provers, it seems natural to ask for efficient-prover protocols for a quantum problem: rather than a classical complexity class like $\AM$, can we achieve such protocols for $\QMA$? This is the ``quantum games PCP'' conjecture, first proposed by Fitzsimons and Vidick ~\cite{10.1145/2688073.2688094}. In 2018, Natarajan and Vidick claimed~\cite{errata-games-qpcp} a resolution of this conjecture; however, this result relied on an earlier result of Vidick~\cite{Vidick2013} on the quantum soundness of the plane-vs-point low-degree test, whose proof turned out to have a bug. Nevertheless, the authors believed that a version of the result with weakened parameters---polylogarithmic instead of logarithmic questions---would hold using the weakened replacement for Vidick's low-degree test result obtained by Ji et al.~\cite{2001.04383}. Unfortunately, it was discovered that the Natarajan and Vidick protocol for $\QMA$ is flawed in \emph{another} way, so that it is now open whether $\QMA$ can be put in $\MIP^*[q = O(\log n), a = O(\log n)]$ with efficient provers even if Vidick's original low-degree result is recovered. In this section we explain Natarajan and Vidick's approach, why it fails, and state a corrected version of their faulty amplification lemma.

\subsection{A template for $\MIP^*$ protocols for $\QMA$}

The protocol in~\cite{errata-games-qpcp} arose from a line of work originated by Fitzsimons and Vidick~\cite{10.1145/2688073.2688094} in 2014. This work gave a prover-efficient protocol for the local Hamiltonian problem with short \emph{quantum} messages, and subconstant soundness gap. The key idea was to distribute the ground state $\ket{\psi}$ amongst multiple provers using an error correcting code as a quantum ``secret sharing'' scheme. Specifically, in their protocol, there are five provers, and the ground state is encoded qubit-by-qubit with the $[[5,1,3]]$ error correcting code with a share being sent to each prover. To verify the energy, the verifier asks each prover for a small number of qubits from their share, and then jointly decodes the shares to measure a single local term of the Hamiltonian on the decoded state.

A series of works~\cite{Zhengfeng,NV16} improved this protocol to one with classical messages, by using \emph{self-testing}, a powerful tool in the nonlocal games literature, to force the provers to perform the measurements themselves on their shares, and honestly report their measurement outcomes to the verifier. The main new technical result of~\cite{errata-games-qpcp}, which now appears in a streamlined and self-contained form in the Appendix of~\cite{2001.04383} is the \emph{quantum low-degree test}, a powerful self-test that can force the provers to perform tensor products of Pauli $X$- and $Z$-measurements, and has constant soundness gap and polylogarithmic question size. 

Using this result, \cite{errata-games-qpcp} gave a ``gap-preseving'' protocol to approximate the ground energy of a Hamiltonian $\H$ consisting of a sum of (possibly high-weight) tensor products of Pauli $X$- and $Z$-operators.
\begin{enumerate}
    \item Ask the provers to share the ground state using a qubit-by-qubit encoding, but using the $[[7,1,3]]$ Steane code. This means there are 7 provers.
    \item Check that the provers share a valid code state, by commanding them to measure the stabilizers of the code. Since the Steane code is CSS, the stabilizers consist only of tensor products of $X$ and $Z$, and so can be measured by the self-test.
    \item Pick a term from the Hamiltonian $\H$. Measure the energy of the provers' state with respect to this term by asking them to measure the corresponding logical operator on the code state. Again, since the Steane code is CSS, and each term of $\H$ is a tensor product of Paulis, the logical operator is itself a tensor product of $X$- and $Z$-operators, and can be measured by the self-test.
\end{enumerate}

This protocol is ``gap-preserving'' in that the optimal success probability of the provers is related linearly to the ground energy of the Hamiltonian independently of the number of qubits: an energy gap of $\delta$ in the Hamiltonian corresponds to a gap in acceptance probabilities of $O(\delta)$, independent of the number of qubits $n$.

In order to use this protocol to solve a $\QMA$-complete problem with a constant soundness gap, it suffices to find a family of Hamiltonians $\H$ of the described form, for which finding the ground energy up to a constant factor is $\QMA$-hard. Natrajan and Vidick attempted this by designing an amplification procedure for the known $\QMA$-hard problem of approximating the ground energy of a local $XX + ZZ$ Hamiltonian up to inverse polynomial gap~\cite{XY-Ham-are-QMA}. 

Unfortunately, this amplification procedure was incorrect\footnote{The issue was first realized by Alex Lombardi who informed Natarajan and Vidick of it in a personal communication.}! The issue arises from the \emph{normalization} of the energy of the Hamiltonian. Typically, one measures the energy on a scale set by the operator norm of $\H$, so that a ``constant energy gap'' means that the energy gap scales as $\Omega(\|\H\|)$ independently of the number of qubits $n$. However, a careful examination of the protocol described above reveals that the relevant scale is not the operator norm, but the $1$-norm of the vector of coefficients of $\H$ in the Pauli basis, a quantity which we define below as the Pauli $1$-norm. While the procedure of \cite{errata-games-qpcp} preserves the operator norm, it causes the Pauli $1$-norm to grow exponentially, destroying the amplified energy gap upon renormalization. In the remainder of this section, we explain why this is the correct normalization to consider, and then describe the Natarajan and Vidick amplification procedure and give the correct scaling of the norms.

\subsection{Measuring the energy of a state via Pauli tensor measurements}

As we described in our summary of Natarajan and Vidick \cite{errata-games-qpcp}, it is a general technique in nonlocal games to be able to force the provers to measure the energy of the specific state $\ket{\psi}$ they possess with respect to \emph{a single} Pauli observable of the verifier's choice. The intention is to use this ability to estimate $\ev{\H}{\psi}$ with the honest provers running the protocol with $\ket{\psi}$ being the state of maximal eigenvalue. Formally, consider a Hamiltonian $\H$ expressed in a decomposition as a summation over Paulis,
\begin{equation} \H = \sum_{P \in S} \beta_P P. \label{eq:decomp-sum-Paulis}\end{equation}
Then the following procedure has an acceptance probability whose bias away from $1/2$ is proportional to the energy of the state.
\begin{mdframed}
\begin{enumerate}
    \item Pick a Pauli $P$ from $S$ with probability $\Pr[P] \defeq \displaystyle \frac{|\beta_P|}{\sum_{P \in S} | \beta_P|}.$
    \item Measure $P$ to get a bit $b \in \{\pm 1\}$.
    \item Accept if $b = \mathrm{sign}(\beta_P)$. 
\end{enumerate}
\end{mdframed}
The acceptance probability can be calculated directly as
\begin{xalign}[eq:Paul-1-norm-of-H]
\Pr[\textrm{accept}] &= \sum_{P \in S} \Pr[P] \cdot \Pr[b = \mathrm{sign}(\beta_P)] \\
&= \sum_{P \in S} \frac{|\beta_P|}{\sum_{P\in S} |\beta_P|} \Pr[ b \cdot \mathrm{sign}(\beta_P)  = 1] \\
&= \sum_{P \in S} \frac{|\beta_P|}{\sum_{P\in S} |\beta_P|} \cdot \left(\frac{1}{2} + \frac{1}{2} \mathrm{sign}(\beta_P) \ev{b}_\psi \right) \\
&= \frac{1}{2} + \frac{1}{2} \cdot \sum_P \frac{\beta_P}{\sum_{P \in S} |\beta_P|} \ev{P}{\psi} \\
&= \frac{1}{2} + \frac{\ev{\H}{\psi} }{2\| \H \|_{\Pp,1}}.
\end{xalign}
In the last line we introduce the notion of the $\norm{\cdot}_{\Pp,1}$, the Pauli 1-norm, which is the minimum weight of $\sum_{P \in S} \abs{\beta_P}$ when a Hamiltonian $\H$ is expressed in a decomposition as a summation over Paulis (as in \eqref{eq:decomp-sum-Paulis}).

Therefore, it is important to note that in \emph{all known} constructions \cite{10.1145/2688073.2688094,errata-games-qpcp} of nonlocal games from local Hamiltonians -- including the proposal by Natarajan and Vidick \cite{errata-games-qpcp} -- rely on a Hamiltonian energy measurement test with analysis analogous to the above calculation. The consequence of the success probability in \eqref{eq:Paul-1-norm-of-H}, is that the promise gap of the resultant nonlocal game family $\Ll_{\text{game}}$ is 
\begin{equation}
\pgap(\Ll_{\text{game}}) \geq \max_{\H \in \Ll_\text{Hamiltonian}} \qty{\frac{1}{\norm{\H}_{\Pp,1}}} \cdot \pgap(\Ll_{\text{Hamiltonian}}).
\end{equation}

Previously, we stated that the Natrajan and Vidick transformation \cite{errata-games-qpcp} generates a reduction from local Hamiltonians to nonlocal games with a promise gap of the game polynomial in the promise gap of the Hamiltonian. This is because they consider $XX+ZZ$ Hamiltonians which by definition have a Pauli 1-norm of at most $O(n^2)$. This is not the case for general Hamiltonians as the Pauli 1-norm can vastly vary from more ``standard'' norms as illustrated by the following example.
\begin{remark}
    The Pauli 1-norm is always an upper bound on the operator norm, but it can be much larger than it. Consider the Hadamard matrix
    \begin{equation} H = \frac{1}{\sqrt{2}} (X + Z).\end{equation}
    This has $\|H\| = 1$ but $\|H\|_{\Pp,1} = \sqrt{2}$. This gap can be exponentially amplified by tensor powers:
    \begin{equation} \| H^{\otimes n} \| = 1, \|H^{\otimes n} \|_{\Pp,1} = \sqrt{2^n}. \end{equation}
\end{remark}
The principal error in Natarajan and Vidick \cite{errata-games-qpcp} is that they considered generating a nonlocal game for the tensor-product amplification of the Hamiltonian $\H$ and they incorrectly calculated the amplification of the Pauli 1-norm in this transformation. Therefore, they incorrectly concluded a nonlocal game of constant promise gap from the tensor-product amplification of a $\QMA$-complete family of $XX+ZZ$ Hamiltonians. The following lemma provides the rectified amplification and accounts for the amplification of the Pauli 1-norm.

\subsection{Hamiltonian promise gap and Pauli 1-norm amplification}

\begin{lemma}[Hamiltonian promise gap amplification]
    Consider a $n$-qudit $\ell$-local Hamiltonian $\H$ such that $-1 \preccurlyeq \H \preccurlyeq 1$ and further assume a promise that $\lambda_{\max}(\H) \geq 1 - 1/p$ or $\lambda_{\max}(\H) \leq 1 - 1/q$. Then there exists an efficient transformation producing a $\ell \cdot k$-local Hamiltonian $\H'$ such that $-1 \preccurlyeq \H' \preccurlyeq 1$ and that 
    \begin{equation}\lambda_{\max}(\H') \geq 1 - \frac{k}{p} \quad \text{or} \quad \lambda_{\max}(\H') \leq 2 \e^{-\frac{k}{2q}} - 1,
    \end{equation}
    respective to the two promised cases. Furthermore, the Pauli 1-norm of $\H'$ can be bounded as
    \begin{equation}
        \| \H' \|_{\Pp,1} \leq 1 + 2 \qty(\frac{1 + \|\H\|_{\Pp,1}}{2})^k.
    \end{equation}
\label{lem:true-amplifcation} \end{lemma}

Note that if $\norm{\H}_{\Pp,1} \leq 1$, then $\norm{\H'}_{\Pp,1} \leq 3$ and is, therefore, bounded. However, it is unknown whether there exists a family of local Hamiltonians which capture $\QMA$ for which the Pauli 1-norm is bounded and amplification is possible. 

One might consider two strategies to adjust a Hamiltonian such that its Pauli 1-norm is manageable. However, neither of these will prove successful for achieving a nonlocal game with consant promise gap. The first would be to be to scale the Hamiltonian $\H \mapsto \theta \H$; for an appropriate $\theta \leq 1/{\norm{\H}_{\Pp,1}}$, this would certainly give a threshold on the Pauli 1-norm but the promise gap parameters of $1 - 1/p$ and $1 - 1/q$ would also move and make amplification difficult. This does not seem like a viable strategy.

The second is to consider applying a randomized restriction to the amplified Hamiltonian $\H'$ to reduce the Pauli 1-norm. Similar to Lemma 36 of \cite{anshu_et_al:LIPIcs.ITCS.2022.6}, we can consider sampling $m = \Omega(n/\delta^2)$ terms from the Hamiltonian $\H'$ to generate a Hamiltonian $\H''$. Application of the operator Chernoff bound \cite[Lemma 2.8]{tropp} gives that 
\begin{equation}
    \Pr\qty[\norm{\H' - \H''} \geq \delta] \leq 2^n \e^{-m \delta^2/32} \leq 1/3.
\end{equation}
If we apply this after having amplified $\H'$ to a constant promise gap, we can select $\delta = O(1)$, we arrive at a Hamiltonian with manifestly fewer terms but a Pauli 1-norm that still scales with $O(n)$, at best. Therefore, the resulting nonlocal games promise gap will not be a constant.

\begin{proof}[Proof of \Cref{lem:true-amplifcation}]
The intention is to amplify the promise gap by considering a tensor product of the original Hamiltonian. We exploit the following two basic facts:
\begin{enumerate}
    \item If $0 \preccurlyeq \mathbf{M}$, then $\lambda_{\max}(\mathbf{M}^{\otimes k}) = \lambda_{\max}(\mathbf{M})^k$.
    \item For small $\eps$, $(1 - \eps)^n \approx 1 - k\eps$.
\end{enumerate}
These facts tell us that if $\mathbf{M}$ is a positive operator whose top eigenvalue is either $1$ or $1- \eps$, then the top eigenvalue of $\mathbf{M}^{\otimes k}$ is either $1$ or bounded away from $1$ by a constant, for $k$ on the order of $1/\eps$. 

To apply this idea to our situation, we first have to linearly shift the Hamiltonian so that the Hamiltonian's spectrum is non-negative. Specifically, we will shift the Hamiltonian by a multiple of identity to make it positive, amplify using by taking tensor product (shown above), and then shift back:
\begin{equation}
    \H' \defeq 2 \qty(\frac{\II + \H}{2})^{\otimes k} - \II.
\end{equation}
By construction $\H'$ has operator norm $\|\H'\| \leq 1$. It remains to bound the its top eigenvalue in the two cases, and to bound its Pauli $1$-norm. Let $\lambda_{\max}(\H) = 1 - 1/a$. Then, %
\begin{align}
    \lambda_{\max}(\H') &= 2 \qty(\frac{1 + \lambda_{\max}(\H)}{2})^k - 1 = 2 \qty(1 - \frac{1}{2a})^k - 1 = 2 \qty(1 - \frac{1}{2a})^{2a \cdot \frac{k}{2a}} - 1.
\end{align}
Well-known bounds can be applied to lower- and upper-bound the top eigenvalue $\lambda_{\max}(\H')$.
\begin{equation}
    2 \qty(1 - \frac{k}{2a}) - 1\leq \lambda_{\max}(\H') \leq 2 \e^{-\frac{k}{2a}} - 1
\end{equation}
Therefore, the promise gap of $1/q - 1/p$ is amplified to at least
\begin{align}
    &\geq 2 \qty(\qty(1 - \frac{k}{2p}) - \e^{-\frac{k}{2q}}) \geq k \qty(\frac{1}{2q} - \frac{1}{p})
\end{align}
This is the ``absolute'' energy gap, prior to normalization by the Pauli $1$-norm\footnote{For the $\QMA$-complete Feynman-Kitaev circuit-to-Hamiltonian construction, the promise cases are constructed with $1/p$ as inverse exponentially small so for $k = \poly(n)$, this amplification is roughly from $1/q$ to $e^{-k/2q}$. Picking $k = 2q$, gives us an amplification of $1/q$ to $1 - 1/e \approx 0.63$.}.
Next we compute the Pauli basis decomposition of $\H'$. Let $S$ be the set of Pauli terms in $H$ and let $m = |S|$.
\begin{xalign}
    \H & = \Exp_{P  \in S} [ \alpha_P P] \\
    \frac{\II + \H}{2} &= \frac{1}{2}\II + \frac{1}{m} \sum_{P \in S} \frac{1}{2} \alpha_P P  \\
    &= \frac{1}{2}\qty(1 + \frac{\alpha_I}{m}) \II + \frac{1}{m} \sum_{P \in S \setminus \{I\}} \frac{1}{2} \alpha_P P \\
    \| (I + \H)/2 \|_{\Pp,1} &\leq \frac{1}{2} + \frac{1}{2} \|\H \|_{\Pp,1} \\
    \| ((I + \H)/2)^{\otimes k} \|_{\Pp,1} &\leq \frac{1}{2^k} (1 + \|\H\|_{\Pp,1})^k \\
    \| 2((I + \H)/2)^{\otimes k} \|_{\Pp,1} &\leq \frac{1}{2^{k-1}} (1 + \|\H\|_{\Pp,1})^k \\
    \| \H' \|_{\Pp,1} &\leq 1 + 2 \qty(\frac{1 + \|\H\|_{\Pp,1}}{2})^k.
\end{xalign}
\end{proof}

\section{Open problems}

Aside from the obvious open conjecture of the quantum games PCP, the following are what we believe to be more accessible stepping stones.

\begin{enumerate}
    \item Is there a $\QMA$-complete family of Hamiltonians with bounded $\| \H\|_{\Pp,1}$ for which \Cref{lem:true-amplifcation} can be applied? This is an interesting question even if the Hamiltonian family does not have the $XX+ZZ$ form that makes it immediately amenable to a nonlocal game reduction.
    \item Towards a negative answer to the previous question, for what classes of  nonlocal Hamiltonians with $\|\H\|_{\Pp,1} \leq 1$  can we show that finding the ground energy up to constant gap (relative to $\|H\|$) is \emph{not} $\QMA$-hard? So far, work on approximations for Hamiltonians has focused on the local case, where ansatzes such as product states are useful. Here we do not expect product states to serve as good ansatzes here, but perhaps more algebraic techniques (e.g. ncSoS) will yield fruitful results. This could be used to show that the local Hamiltonian problem is not as hard as it seems.
    \item More sophisticated gap amplification schemes have been studied \cite{hastings2007area_law,AradLV12,AradKLV13,AHS20,2d-arealaw, anshu_et_al:LIPIcs.ITCS.2022.6,anshu2022nlts} that involve applying some polynomial $f$ to $\H$. How can we bound $\|f(\H)\|_{\Pp,1}$, in terms of $\|\H\|_{\Pp,1}$ and properties of $f$?
    \item Is there a template for a 2-prover version of Natarajan-Vidick \cite{errata-games-qpcp}? That is, assuming any favorable conjecture in Hamiltonian complexity, can one find a two-prover succinct $\MIP^*$ for $\QMA$, where the provers are efficient given copies of the ground state?
\end{enumerate}
\label{sec:errata}

\section{Acknowledgements}
We acknowledge Alex Lombardi for finding the bug in \cite{DBLP:conf/focs/NatarajanV18}, as well as Tony Metger, Thomas Vidick, and Tina Zhang for many helpful discussions and sharing an unpublished result of theirs. We also thank Thomas Vidick for detailed comments on an early draft of this article. This work was partially completed while AN and CN were participants in the Simons Institute for the Theory of Computing workshop on \emph{Quantum Algorithms, Complexity, and Fault Tolerance}.

\bibliography{references}
\bibliographystyle{myhalpha}

\end{document}